\documentclass[aps,prl,twocolumn,showpacs,byrevtex]{revtex4}
\usepackage{graphicx}    

\begin{document}
\title{Quantum Phase Transition in an Antiferromagnetic Spinor Bose-Einstein Condensate}
\author{E. M. Bookjans}
\author{A. Vinit}
\author{C. Raman}
\email{chandra.raman@physics.gatech.edu}
\affiliation{School of Physics, Georgia Institute of Technology, Atlanta, Georgia 30332, USA}
\date{\today}

\begin{abstract}

We have experimentally observed the dynamics of an antiferromagnetic sodium Bose-Einstein condensate (BEC) quenched through a quantum phase transition.  Using an off-resonant microwave field coupling the $F = 1$ and $F = 2$ atomic hyperfine levels, we rapidly switched the quadratic energy shift $q$ from positive to negative values.  At $q = 0$ the system undergoes a transition from a polar to antiferromagnetic phase.  We measured the dynamical evolution of the population in the $F = 1, m_F = 0$ state in the vicinity of this transition point and observed a mixed state of all 3 hyperfine components for $q < 0$.  We also observed the coarsening dynamics of the instability for $q<0$, as it nucleated small domains that grew to the axial size of the cloud. 
\end{abstract}

\maketitle

A quantum phase transition describes a many-body system whose ground state can be tuned through a point of non-analyticity \cite{sachdev99}.  Quantum gases afford the possibility to realize such phase transitions in the laboratory, as well as to explore the dynamical evolution of the state of the system by directly controlling the tuning parameters.  In particular, spinor Bose-Einstein condensates possess a vector order parameter with additional degrees of freedom relevant to this problem.  By changing external fields dynamically, one can observe and quantify a host of non-equilibrium phenomena, including spin domain formation and aggregation, topological defect creation, and possible dynamic scaling laws \cite{ueda2010}. 

In this work we examine a first order phase transition associated with the quadratic energy shift, which is an essential parameter in spinor physics.  For a spin-1 BEC the spin-dependent Hamiltonian can be written in a mean-field representation as 
\[ H_{sp} = \frac{c_2}{2} n \langle {\bf \hat{F}} \rangle^2 + q \langle \hat{F}_z ^2 \rangle \]
where ${\bf \hat{F}},\hat{F}_z$ are the vector spin-1 operator and its $z$-projection, respectively, $n$ is the particle density, $c_2$ the spin-dependent interaction coefficient \footnote{In terms of atomic parameters, $c_2 = \frac{4 \pi \hbar^2}{3 M}(a_2-a_0)$, where $M$ is the atomic mass, and $a_{2,0}$ are the triplet and singlet scattering lengths, respectively.}, and $q$ is the energy difference $\frac{1}{2}(E_{+1}+E_{-1})-E_0$, where $E_i$ is the energy of the atomic level for the spin $m_F = i$ component of $F = 1$. The spin-dependent interaction coefficient $c_2$ arises from spin-changing collisions that can convert two $m_F= 0$ atoms into an $m_F = \pm 1$ pair and vice-versa, a process constrained by the conservation of angular momentum.  It determines the nature of the ground state--antiferromagnetic for $c_2>0$ or ferromagnetic for $c_2<0$. The quadratic energy shift $q$ is usually determined by an external magnetic field $B$ through the second-order Zeeman effect, and is $\propto B^2$.  However, 
it can also be tuned using a microwave dressing field \cite{gerbier2006,leslie2009}, a feature we exploit in this work to uncover a previously unexplored phase transition in an antiferromagnetic $^{23}$Na spinor gas.  While in general the levels shift independently, spin conservation leads to the cancellation of the linear energy shifts, such that only the quadratic energy shift $q$ is important for spinor BECs.  Therefore $q$ plays the role of an external parameter and the combination of $c_2$ and $q$ realizes a rich phase diagram of possibilities \cite{zhang03}.  Various quantum phases and dynamics have been observed for both $c_2 > 0$ and $c_2 < 0$ \cite{sten98spin,chan04,sadler06,black07,liu09,klempt2009,kronjager2010}. 

For an antiferromagnetic spinor BEC constrained to have zero net magnetization, the ground state solution is a nematic order parameter $\Psi$.  It varies smoothly with $q$ for all values except $q = 0$, which divides the phase diagram into two regions.  For $q > 0$ the ground state is a polar condensate consisting of a single component--the $m_F = 0$ spin projection that minimizes $\langle \hat{F}_z ^2 \rangle$.  For $q<0$ the ground state maximizes the same quantity through a superposition of two components $m_F = \pm 1$, a so-called antiferromagnetic phase \cite{ueda2010}.  The symmetry properties of the ground state therefore change discontinuously, defining a first order phase boundary.  

Exactly at the phase transition point the many-body ground state is a condensate of boson pairs forming a spin singlet state, and possessing super-Poissonian spin fluctuations.  Near this boundary the mean-field wavefunction $\Psi$ undergoes collapse and revival dynamics triggered by quantum fluctuations \cite{law98spin2,cui2008,barnett2010}.  Controllably accessing $q = 0$ would restore the full $S^2$ symmetry of the nematic order parameter, which has unusual topological defects such as half-quantum vortices \cite{zhou01, wright2009,ruostekoski2003}.  However, experimentally this requires low magnetic fields where spinor condensates are susceptible to ambient magnetic field noise that can mask interaction-related phenomena.  In this letter, we investigated the dynamical instability of an $F=1, m_F = 0$ antiferromagnetic sodium BEC that is rapidly quenched across the boundary from $q>0$ to $q<0$.  The quadratic energy shift $q$ was tuned by an additional microwave dressing field that allowed us to access the $q<0$ regime.


Microwave dressing of ferromagnetic $^{87}$Rb has been used to tune spin mixing dynamics into the resonant regime in optical lattices \cite{gerbier2006}, and for the study of spontaneous magnetization \cite{leslie2009}.  Unlike ferromagnetic systems, in antiferromagnetic condensates one generally expects unmagnetized domains to form as a result of $m_F=\pm 1$ pair production.  Pair formation dynamics have also been studied in antiferromagnetic $F = 2$ $^{87}$Rb spinors where the sign of both $c_2$ and $q$ are reversed with respect to the $F = 1$ manifold \cite{schmaljohann04,klempt2009,klempt2010}, although the spinor lifetime is limited by hyperfine state changing collisions. The $F = 1$ spin system, by contrast, is intrinsically stable and amenable to studies of long time scale dynamics in the transition region--in the present work we observed evolution times ranging from 30 ms to over 2 seconds.

We prepared BECs with up to $3 \times 10^{6}$ sodium atoms with a peak atom density $n_0 = 5.4 \times 10^{14}$ cm$^{-3}$  in an optical dipole trap created by a single focused far-detuned 1064 nm laser beam. The measured axial trapping frequency was 8 Hz, with inferred radial frequencies of about 600 Hz, which correspond to Thomas-Fermi radii of $4~\mu$m and $270~\mu$m, respectively. The condensate is initially prepared in a magnetic trap from which it was transferred into the optical trap as described in earlier work \cite{naik05}.  In order to create a pure $m_F =0$ condensate, we adiabatically swept the frequency of an rf magnetic field across the $m_F =-1 \rightarrow m_F = 0$ transition at a bias magnetic field of 13 Gauss \cite{mewe97}. The bias was then ramped to a final value of $B = 97$~mG in 30~ms, which was defined as the starting point (t = 0) of our experiment. 

At this magnetic field, the quadratic Zeeman shift is $q_B = h \times$ 2.5 Hz \cite{sten98spin}, far smaller than the spin-dependent interaction energy $c_2 n_0  = h \times$130 Hz \footnote{The peak atom density $n_0$ was determined from the time-of-flight expansion velocity.  We used the value $c_2 = 1.6 \times 10^{-52}$~J-m$^3$ \protect{\cite{black07}}.}. The radial Thomas-Fermi radius of our cloud, 4$\mu$m, is only a factor of 3 larger than the spin-healing length $\xi_s = \hbar / \sqrt{2Mc_2n} \approx 1.3$~$\mu$m, which is the width of a typical spin domain wall \cite{stam99thesis}.  Therefore our experiment is performed mostly in a quasi-1D geometry, in which only longitudinal spin structures are expected.

\begin{figure}[htp]
\centering
\includegraphics[width=\columnwidth]{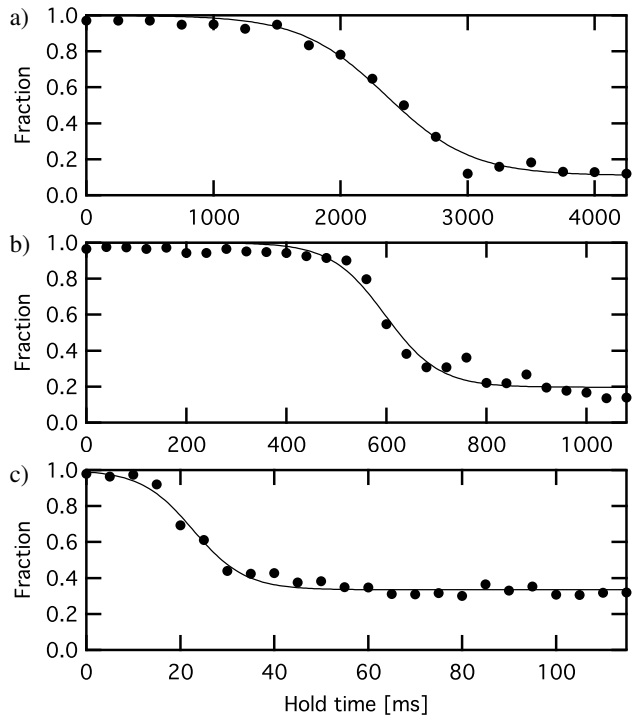}
\vspace{-0.27 in}
\caption{Quenching dynamics.  For different final values of $q$, the $m_F = 0$ fraction (circles) is plotted versus time after the quench.  The solid lines are fits of the $m_F =0$ fraction to a Sigmoid function. In a) no microwave dressing field is applied.  This data shows the relaxation in the absence of a quench ($q =h\times 2.5$~Hz).  The equilibration to a pure $m_F = \pm1$ cloud occurs after 2.1~s. In b) the gas is quenched to $q=h\times -3.2$~Hz, showing that the population decay is faster. In c) the gas is quenched to  $q=h\times -17.4$~Hz and rapidly reaches a quasi-equilibrium state containing all 3 spin components in roughly equal proportions, i.e. an order parameter delocalized over the S$^2$ sphere.  Each data point in the figure corresponded to a separate run of the experiment.}
\label{fig:dynamics}
\vspace{-0.25in}
\end{figure}

At $t=0$ we instantly turned on an oscillating microwave field within tens of $\mu$s, far shorter than any dynamical timescale relevant to the problem.  We used microwave powers between 0 and 7.5 Watts, which allowed us to tune $q$ from its initial value of $q_B$ to a final value $q = q_B + q_M = $ -18.5~Hz, where $q_M$ is the quadratic energy shift due to the AC Stark shift caused by the microwave field \cite{gerbier2006}.  This field created an instability in the initial $m_F = 0$ spin state.  Holding the condensate at a fixed value of $q$ and varying the hold time after the quench was initiated, we could observe the temporal evolution of the fractional population in $m_F = 0$.  This corresponded to a measurement of the squared amplitude of the z-component of the nematic order parameter.  The relative populations in the $m_F = -1, 0, +1$ states were determined by Stern-Gerlach time-of-flight images. After an expansion of 3-3.5~ms, we pulsed on a Stern-Gerlach field gradient for a duration of 2-4.5 ms perpendicular to the axial direction separating the 3 components spatially \cite{sten98spin}.  After a total time-of-flight of 25 ms, we optically pumped the atoms into the $F = 2$ state and imaged them on the $F= 2 \rightarrow F' = 3 $ cycling transition ensuring an equal imaging sensitivity to each spin component.  Examples of the temporal evolution for $q = h \times -3.2$~Hz and $q = h \times -17.4$~Hz are shown in Figures \ref{fig:dynamics}b) and c), respectively.  They indicate that a pure $m=0$ condensate was unstable, evolving into a superposition of all 3 components that preserved the overall zero net magnetization.  In order to quantify the $q$-dependence of the instability, we fitted data such as shown in Figure \ref{fig:dynamics} to a Sigmoid function and determined both the cross-over time $T_{1/2}$ and the final saturation value $f_{0,min}$ where $f_0(T_{1/2}) = 1/2(1+f_{0,min})$.  We defined the measured instability rate to be $\Gamma (q) \equiv 1/T_{1/2}$.

At zero microwave power, the quadratic energy shift was $q_B/(c_2 n_0) = 0.02$.  Under these conditions, the gas was relatively stable against spin relaxation, i.e. the creation of $m_F = \pm1$ pairs.  This stability is a characteristic of quantum antiferromagnetism \cite{ueda2010}.  For a homogeneous system and $0 < q < c_2 n$ the $m_F = 0$ state is stabilized against the creation of magnon excitations (spin waves) by an energy gap $\Delta = \sqrt{(\epsilon_k + q)(\epsilon_k+q+2 c_2 n)}$, where $\epsilon_k \equiv \frac{\hbar^2 k^2}{2 M}$ \cite{saito05}.  The $k=0$ mode is the most unstable mode and for $q<<c_2n$, the energy gap $\approx \sqrt{2 q c_2 n} = h \times $25 Hz for our parameters.  

As $q \rightarrow 0$, quantum fluctuations destabilize the pure $m_F = 0$ state:  the fraction of atoms in the $m_F =\pm 1$ states should reach of order 1 within a time $\sim 1.5$ seconds \cite{cui2008}. This time scale is consistent with the slow rate of relaxation ($\sim 0.5$ s$^{-1}$) that we observed in our experiment (see Figure \ref{fig:dynamics} a)). However, we cannot definitively rule out other mechanisms, including thermal fluctuations (our BEC had a thermal fraction of 40\%) and imperfect transfer to the $m_F=0$ state.  For hold times longer than 2 seconds the cloud separated into two non-overlapping $m_F = \pm 1$ spin domains along the long axis of our trap, an indication that small residual magnetic field gradients might have been present.

For $-c_2 n < q < 0$ the gas becomes unstable against pair formation due to the presence of an imaginary frequency $\Gamma = \mathcal{I}m(\Delta)$.  As $|q|$ increases, the gas progressively evolves into a mixed state.  In all cases the temporal dependence of the $m_F = 0$ fraction, $f_0 (t)$, followed a backward $S$-shaped curve that saturated at a value $f_{0,min}$ depending on $q$.  The data sets could be roughly divided into 2 regions:  for positive (Figure \ref{fig:dynamics}a)) and slightly negative $q$ (Figure \ref{fig:dynamics}b)) we observed a slow decay to a value $f_{0,min}$ nearly zero.  For this data, an examination of the spatial distribution of the three hyperfine components showed that the $\pm 1$ states had separated from one another along the axial direction.  This is most likely due to the residual magnetic field gradients mentioned earlier.  

For more negative $q$ (Figure \ref{fig:dynamics}c)), however, the behavior was dramatically different--$f_0$ approached a final value of $f_{0,min} \approx 0.3$ within a time as little as 30 ms and remained roughly constant over 200 ms \footnote{We could not observe the gas for longer times due to losses caused by residual excitation to the $F=2$ manifold at the microwave power and detuning used.}.  Thus for the range of $q$ explored in this work, the instability was observed to create a mixed state of all 3 components which appeared to be metastable on a timescale much longer than $\Gamma^{-1}$.  When accounting for residual thermal atoms, the estimated condensate fraction $m_F = 0$ is slightly less than $0.3$.

\begin{figure}[htp]
\centering
\includegraphics[width=\columnwidth]{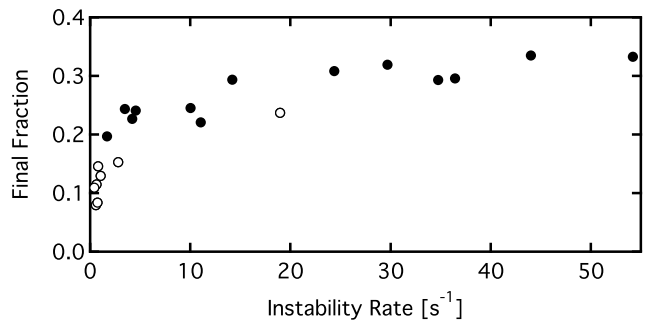}
\vspace{-0.27 in}
\caption{Final $m_F = 0$ fraction versus instability rate $\Gamma$. The open circles indicate data for which the final $m_F=0$ distribution consisted purely of thermal atoms.}
\label{fig:satval}
\end{figure}

Figure \ref{fig:satval} shows the final saturation value $f_{0,min}(q)$ plotted versus $\Gamma(q)$.  These data provide further evidence for the two regimes mentioned earlier.  Above a critical instability rate of $\simeq $ 3 s$^{-1}$, the final fraction $ f_{0,min}$ was between 0.2 and 0.3, and the $m_F = 0$ state maintained a significant presence in the cloud.  With the exception of one data point, only for the very lowest instability rates $< $ 3  s$^{-1}$ was the steady state Bose-condensed spin distribution consistent with a pure $m_F = \pm 1$ spin mixture (open circles).  

We can qualitatively understand the instability in terms of a quantum rotor model that is valid in the single mode approximation \cite{barnett2010}.  Such a model cannot describe spin domain formation, and therefore is inapplicable to the data for small $|q|$ discussed above, for which we observed spin segregation.  However, it may provide useful insights into the short timescale behavior following the quench for larger $|q|$, where a mixture of all 3 components was observed.  The system is described by a single, macroscopic quantum rotor with angular momentum $L$, moment of inertia $I = N \hbar^2/(c_2 n)$, and a potential energy term $ V\approx q N sin^2(\theta)$ for our parameters.  $\theta = 0$ describes a pure $m=0$ state, while $0 < \theta < \pi$ corresponds to the inclusion of $\pm 1$ pairs into the wavefunction.  The sudden quench transition from $q>0$ to $q<0$ causes $V$ to change sign, resulting in rapid dispersion of a wavepacket initially highly localized in angle near $\theta=0$.  The resulting wavepacket dynamics are mostly classical in character and consist of a rapid dispersion in $\theta$ followed by sparse, periodic revivals at time $t_{rev}$ \cite{stapelfeldt2003} .  For our parameters $t_{rev} \sim 350$ seconds, considerably longer than our observation time \cite{barnett2010}.  For short times, as in Figure \ref{fig:dynamics}c), we observed only the rapid dispersion phase, and we interpret the measured value of $\approx 0.3$ for each spin component to be the result of wavepacket dispersion that tends to equalize the spin populations.

\begin{figure}[htbp]

\centering
\includegraphics[width = \columnwidth]{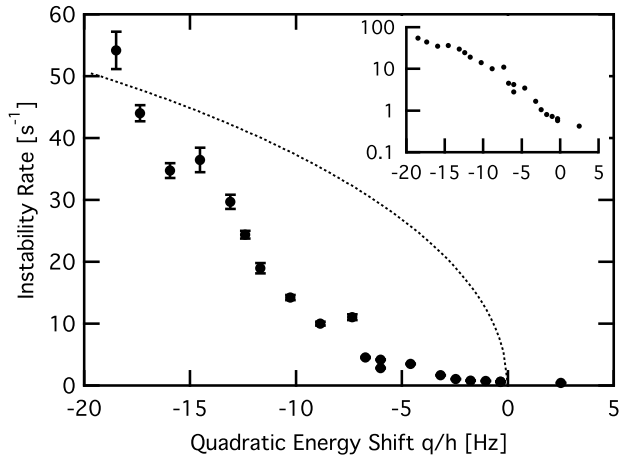}
\vspace{-0.27 in}
\caption{Quenching through the quantum phase transition.  The formation rate of $m_F=\pm 1$ atom pairs is plotted versus the quadratic energy shift $q$ (circles) determined by fitting the temporal evolution of the $m_F =0$ fraction to a Sigmoid function. The error bars give the statistical uncertainty in the fit. The instability rate dramatically increases below the transition point at $q=0$. Also shown (dotted line) is the predicted instability rate from Bogoliubov theory for a uniform gas. Inset shows the same data plotted on a semilog scale.}  \label{fig:critical_point}
\end{figure}

The measured instability rate $\Gamma$ has been plotted versus the final quadratic energy $q = q_B+q_M$ in Figure \ref{fig:critical_point} for $q$ ranging from + 2.5 Hz to -18.5 Hz.  The data shows a steep rise in the pair formation rate by a factor of nearly 100 as $\frac{q}{c_2 n_0}$ varied from $+0.02$  to $-0.15$, indicating that we had crossed a phase boundary in the dynamical evolution of the system.  Figure \ref{fig:critical_point} also shows the predicted maximal instability rate for $-c_2 n_0 < q<0$ from Bogoliubov theory for a uniform $m_F=0$ gas with the same average density $\langle n \rangle$, $\Gamma_{unif} = \sqrt{|q|(q+2 c_2 \langle n \rangle)}$.  This corresponds to the formation of correlated pairs of atoms in a spatial mode with wavevector $k=0$, i.e. a homogeneous rate of pair formation throughout the cloud \cite{saito05}.  The homogeneous theory is in considerable disagreement with our data, which could be attributed to the finite size and 1-D geometry of our trap.  This departure is consistent with earlier work on pair formation dynamics in $F=2$ spinor condensates which highlighted resonant structures and the importance of the inhomogeneous density profile in determining the modes that were populated \cite{klempt2009,scherer2010}. No clear indication of resonances were visible in our data.  Moreover, this theory does not account for possible spin exchange processes between the condensate and the residual thermal cloud, which could play a role in our observations \cite{mcguirk03}.  

\begin{figure}[htbp]
\centering
\includegraphics[width = \columnwidth]{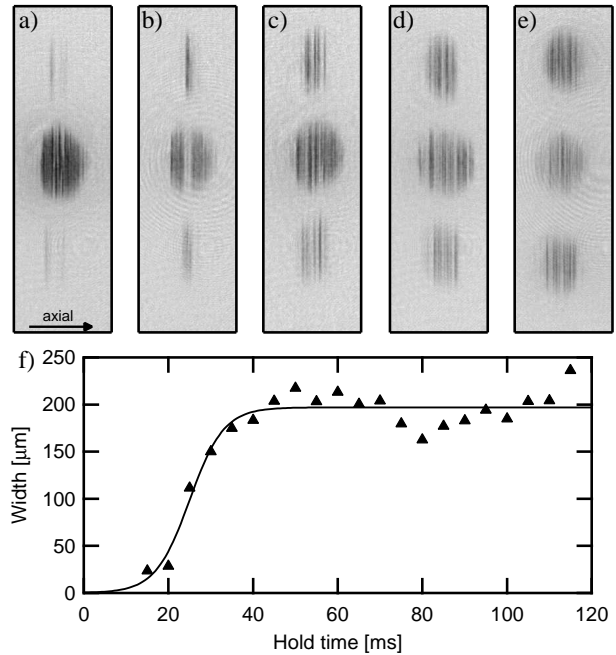}
\vspace{-0.27 in}
\caption{Spatial dynamics of the instability. Absorption images of the condensate taken at a time of flight (TOF) of 25~ms for a quench to $q = \times -17.4$~Hz for different hold times: a) 15~ms, b) 20~ms, c) 25~ms, d) 30~ms, and e) 150~ms. From top to bottom the images show the $m_F = -1$, $m_F = 0$, and $m_F =+1$ spin state distribution. The width of the images is 1~mm. In f) the width of the $m_F = -1$ component after a $\text{TOF}=25$~ms, determined by a Gaussian fit, is plotted as a function of hold time (triangles). The fit to a Sigmoid function (line) is included to guide the eye. The appearance of a small $m_F =  \pm1$ domains, which grow outward with hold time, contradicts a homogenous Bogoliubov theory, which predicts a uniform instability rate across the cloud.}  \label{fig:domains}
\vspace{-0.25 in}
\end{figure}

In order to gain an intuitive understanding of the spatial dynamics one can think of the inhomogeneous $m_F=0$ condensate as a locally varying gain medium for the pair formation instability.  For very short times after the quench, depletion of the gain can be neglected, and one may write the growth rate of $m_F=\pm 1$ atom pair number for $q<0$ using a local density approximation as $ \Gamma_{local} = \sqrt{ |q| (q+2 c_2 n(\vec{r}))} / h $, where $n(\vec{r})$ is the spatial density profile of the $m_F=0$ cloud with the maximum gain occurring at the cloud center.  In this regime, the inhomogeneous gain acts as a nonlinear spatial mode coupler that converts energy from long to short wavelengths, i.e. exhibits a tendency to nucleate small sized domains.  These dynamics are not captured in the homogeneous theory, but appear in our data.  For data sets with $q < h \times$ -7 Hz, the $m_F=\pm 1$ atom distribution appeared initially as one or more small axial domains near the cloud center.  As an example, Figure \ref{fig:domains}a)-e) show Stern-Gerlach images at various times after the quench for $q = h \times -17.4$~Hz.  These domains appeared to coalesce into a larger domain that grew in size with time until it became comparable to the axial Thomas-Fermi radius (see Figure \ref{fig:domains}f)).   

Once a substantial number of atom pairs have been created the $m_F=0$ condensate can be {\em locally} depleted (see for Example \ref{fig:domains}b)).  Since the $m_F=\pm 1$ and $m_F=0$ clouds are immiscible for $c_2 > 0$, this creates a potential well that traps the pairs but allows the domain to grow axially due to the continued effect of the instability (expansion along the radial direction costs a substantial kinetic energy due to the tighter confinement).  Thus the long timescale evolution of the instability exhibits one-dimensional coarsening dynamics \cite{bray94}.  We also observed smaller domain structures which could not be quantified clearly due to the presence of undamped $m_F=0$ density fluctuations in the initial state caused by nonadiabaticity in the initial transfer to the optical trap.  

In conclusion, we have tuned an antiferromagnetic condensate through a phase boundary and quantified in detail the rate of instability in its vicinity. 
Future work will explore the phase coherence between the dynamically created $m_F = \pm 1$ spin components in relation to topological defect formation.  

We thank Carlos Sa de Melo for valuable discussions and Jason Gilbertson, Sean Dixon, and Diego Remolina for technical assistance. This work was supported by the U.S. Department of Energy.
%

\begin{thebibliography}{28}
\expandafter\ifx\csname natexlab\endcsname\relax\def\natexlab#1{#1}\fi
\expandafter\ifx\csname bibnamefont\endcsname\relax
  \def\bibnamefont#1{#1}\fi
\expandafter\ifx\csname bibfnamefont\endcsname\relax
  \def\bibfnamefont#1{#1}\fi
\expandafter\ifx\csname citenamefont\endcsname\relax
  \def\citenamefont#1{#1}\fi
\expandafter\ifx\csname url\endcsname\relax
  \def\url#1{\texttt{#1}}\fi
\expandafter\ifx\csname urlprefix\endcsname\relax\def\urlprefix{URL }\fi
\providecommand{\bibinfo}[2]{#2}
\providecommand{\eprint}[2][]{\url{#2}}

\bibitem[{\citenamefont{Sachdev}(1999)}]{sachdev99}
\bibinfo{author}{\bibfnamefont{S.}~\bibnamefont{Sachdev}},
  \emph{\bibinfo{title}{Quantum phase transitions}}
  (\bibinfo{publisher}{Cambridge University Press}, \bibinfo{address}{Cambridge
  ; New York}, \bibinfo{year}{1999}).

\bibitem[{\citenamefont{Ueda and Kawaguchi}(2010)}]{ueda2010}
\bibinfo{author}{\bibfnamefont{M.}~\bibnamefont{Ueda}} \bibnamefont{and}
  \bibinfo{author}{\bibfnamefont{Y.}~\bibnamefont{Kawaguchi}},
  \bibinfo{journal}{eprint:arXiv.org/abs/1001.2072}  (\bibinfo{year}{2010}).

\bibitem[{\citenamefont{Gerbier et~al.}(2006)\citenamefont{Gerbier, Widera,
  F\"olling, Mandel, and Bloch}}]{gerbier2006}
\bibinfo{author}{\bibfnamefont{F.}~\bibnamefont{Gerbier}},
  \bibinfo{author}{\bibfnamefont{A.}~\bibnamefont{Widera}},
  \bibinfo{author}{\bibfnamefont{S.}~\bibnamefont{F\"olling}},
  \bibinfo{author}{\bibfnamefont{O.}~\bibnamefont{Mandel}}, \bibnamefont{and}
  \bibinfo{author}{\bibfnamefont{I.}~\bibnamefont{Bloch}},
  \bibinfo{journal}{Physical Review A} \textbf{\bibinfo{volume}{73}},
  \bibinfo{pages}{041602} (\bibinfo{year}{2006}).

\bibitem[{\citenamefont{Leslie et~al.}(2009)\citenamefont{Leslie, Guzman,
  Vengalattore, Sau, Cohen, and Stamper-Kurn}}]{leslie2009}
\bibinfo{author}{\bibfnamefont{S.~R.} \bibnamefont{Leslie}},
  \bibinfo{author}{\bibfnamefont{J.}~\bibnamefont{Guzman}},
  \bibinfo{author}{\bibfnamefont{M.}~\bibnamefont{Vengalattore}},
  \bibinfo{author}{\bibfnamefont{J.~D.} \bibnamefont{Sau}},
  \bibinfo{author}{\bibfnamefont{M.~L.} \bibnamefont{Cohen}}, \bibnamefont{and}
  \bibinfo{author}{\bibfnamefont{D.~M.} \bibnamefont{Stamper-Kurn}},
  \bibinfo{journal}{Physical Review A} \textbf{\bibinfo{volume}{79}},
  \bibinfo{pages}{043631} (\bibinfo{year}{2009}).

\bibitem[{\citenamefont{Zhang et~al.}(2003)\citenamefont{Zhang, Yi, and
  You}}]{zhang03}
\bibinfo{author}{\bibfnamefont{W.}~\bibnamefont{Zhang}},
  \bibinfo{author}{\bibfnamefont{S.}~\bibnamefont{Yi}}, \bibnamefont{and}
  \bibinfo{author}{\bibfnamefont{L.}~\bibnamefont{You}}, \bibinfo{journal}{New
  J Phys} \textbf{\bibinfo{volume}{5}}, \bibinfo{pages}{77}
  (\bibinfo{year}{2003}).

\bibitem[{\citenamefont{Stenger et~al.}(1998)\citenamefont{Stenger, Inouye,
  Stamper-Kurn, Miesner, Chikkatur, and Ketterle}}]{sten98spin}
\bibinfo{author}{\bibfnamefont{J.}~\bibnamefont{Stenger}},
  \bibinfo{author}{\bibfnamefont{S.}~\bibnamefont{Inouye}},
  \bibinfo{author}{\bibfnamefont{D.~M.} \bibnamefont{Stamper-Kurn}},
  \bibinfo{author}{\bibfnamefont{H.-J.} \bibnamefont{Miesner}},
  \bibinfo{author}{\bibfnamefont{A.~P.} \bibnamefont{Chikkatur}},
  \bibnamefont{and} \bibinfo{author}{\bibfnamefont{W.}~\bibnamefont{Ketterle}},
  \bibinfo{journal}{Nature} \textbf{\bibinfo{volume}{396}},
  \bibinfo{pages}{345} (\bibinfo{year}{1998}).

\bibitem[{\citenamefont{Chang et~al.}(2004)\citenamefont{Chang, Hamley,
  Barrett, Sauer, Fortier, Zhang, You, and Chapman}}]{chan04}
\bibinfo{author}{\bibfnamefont{M.~S.} \bibnamefont{Chang}},
  \bibinfo{author}{\bibfnamefont{C.~D.} \bibnamefont{Hamley}},
  \bibinfo{author}{\bibfnamefont{M.~D.} \bibnamefont{Barrett}},
  \bibinfo{author}{\bibfnamefont{J.~A.} \bibnamefont{Sauer}},
  \bibinfo{author}{\bibfnamefont{K.~M.} \bibnamefont{Fortier}},
  \bibinfo{author}{\bibfnamefont{W.}~\bibnamefont{Zhang}},
  \bibinfo{author}{\bibfnamefont{L.}~\bibnamefont{You}}, \bibnamefont{and}
  \bibinfo{author}{\bibfnamefont{M.~S.} \bibnamefont{Chapman}},
  \bibinfo{journal}{Physical Review Letters} \textbf{\bibinfo{volume}{92}},
  \bibinfo{pages}{140403} (\bibinfo{year}{2004}).

\bibitem[{\citenamefont{Sadler et~al.}(2006)\citenamefont{Sadler, Higbie,
  Leslie, Vengalattore, and Stamper-Kurn}}]{sadler06}
\bibinfo{author}{\bibfnamefont{L.~E.} \bibnamefont{Sadler}},
  \bibinfo{author}{\bibfnamefont{J.~M.} \bibnamefont{Higbie}},
  \bibinfo{author}{\bibfnamefont{S.~R.} \bibnamefont{Leslie}},
  \bibinfo{author}{\bibfnamefont{M.}~\bibnamefont{Vengalattore}},
  \bibnamefont{and} \bibinfo{author}{\bibfnamefont{D.~M.}
  \bibnamefont{Stamper-Kurn}}, \bibinfo{journal}{Nature}
  \textbf{\bibinfo{volume}{443}}, \bibinfo{pages}{312} (\bibinfo{year}{2006}).

\bibitem[{\citenamefont{Black et~al.}(2007)\citenamefont{Black, Gomez, Turner,
  Jung, and Lett}}]{black07}
\bibinfo{author}{\bibfnamefont{A.~T.} \bibnamefont{Black}},
  \bibinfo{author}{\bibfnamefont{E.}~\bibnamefont{Gomez}},
  \bibinfo{author}{\bibfnamefont{L.~D.} \bibnamefont{Turner}},
  \bibinfo{author}{\bibfnamefont{S.}~\bibnamefont{Jung}}, \bibnamefont{and}
  \bibinfo{author}{\bibfnamefont{P.~D.} \bibnamefont{Lett}},
  \bibinfo{journal}{Physical Review Letters} \textbf{\bibinfo{volume}{99}},
  \bibinfo{pages}{070403} (\bibinfo{year}{2007}).

\bibitem[{\citenamefont{Liu et~al.}(2009)\citenamefont{Liu, Jung, Maxwell,
  Turner, Tiesinga, and Lett}}]{liu09}
\bibinfo{author}{\bibfnamefont{Y.}~\bibnamefont{Liu}},
  \bibinfo{author}{\bibfnamefont{S.}~\bibnamefont{Jung}},
  \bibinfo{author}{\bibfnamefont{S.~E.} \bibnamefont{Maxwell}},
  \bibinfo{author}{\bibfnamefont{L.~D.} \bibnamefont{Turner}},
  \bibinfo{author}{\bibfnamefont{E.}~\bibnamefont{Tiesinga}}, \bibnamefont{and}
  \bibinfo{author}{\bibfnamefont{P.~D.} \bibnamefont{Lett}},
  \bibinfo{journal}{Physical Review Letters} \textbf{\bibinfo{volume}{102}},
  \bibinfo{pages}{125301} (\bibinfo{year}{2009}).

\bibitem[{\citenamefont{Klempt et~al.}(2009)\citenamefont{Klempt, Topic,
  Gebreyesus, Scherer, Henninger, Hyllus, Ertmer, Santos, and
  Arlt}}]{klempt2009}
\bibinfo{author}{\bibfnamefont{C.}~\bibnamefont{Klempt}},
  \bibinfo{author}{\bibfnamefont{O.}~\bibnamefont{Topic}},
  \bibinfo{author}{\bibfnamefont{G.}~\bibnamefont{Gebreyesus}},
  \bibinfo{author}{\bibfnamefont{M.}~\bibnamefont{Scherer}},
  \bibinfo{author}{\bibfnamefont{T.}~\bibnamefont{Henninger}},
  \bibinfo{author}{\bibfnamefont{P.}~\bibnamefont{Hyllus}},
  \bibinfo{author}{\bibfnamefont{W.}~\bibnamefont{Ertmer}},
  \bibinfo{author}{\bibfnamefont{L.}~\bibnamefont{Santos}}, \bibnamefont{and}
  \bibinfo{author}{\bibfnamefont{J.~J.} \bibnamefont{Arlt}},
  \bibinfo{journal}{Physical Review Letters} \textbf{\bibinfo{volume}{103}},
  \bibinfo{pages}{195302} (\bibinfo{year}{2009}).

\bibitem[{\citenamefont{Kronjager et~al.}(2010)\citenamefont{Kronjager, Becker,
  Soltan-Panahi, Bongs, and Sengstock}}]{kronjager2010}
\bibinfo{author}{\bibfnamefont{J.}~\bibnamefont{Kronjager}},
  \bibinfo{author}{\bibfnamefont{C.}~\bibnamefont{Becker}},
  \bibinfo{author}{\bibfnamefont{P.}~\bibnamefont{Soltan-Panahi}},
  \bibinfo{author}{\bibfnamefont{K.}~\bibnamefont{Bongs}}, \bibnamefont{and}
  \bibinfo{author}{\bibfnamefont{K.}~\bibnamefont{Sengstock}},
  \bibinfo{journal}{Physical Review Letters} \textbf{\bibinfo{volume}{105}},
  \bibinfo{pages}{090402} (\bibinfo{year}{2010}).

\bibitem[{\citenamefont{Law et~al.}(1998)\citenamefont{Law, Pu, and
  Bigelow}}]{law98spin2}
\bibinfo{author}{\bibfnamefont{C.~K.} \bibnamefont{Law}},
  \bibinfo{author}{\bibfnamefont{H.}~\bibnamefont{Pu}}, \bibnamefont{and}
  \bibinfo{author}{\bibfnamefont{N.~P.} \bibnamefont{Bigelow}},
  \bibinfo{journal}{Physical Review Letters} \textbf{\bibinfo{volume}{81}},
  \bibinfo{pages}{5257} (\bibinfo{year}{1998}).

\bibitem[{\citenamefont{Cui et~al.}(2008)\citenamefont{Cui, Wang, and
  Zhou}}]{cui2008}
\bibinfo{author}{\bibfnamefont{X.}~\bibnamefont{Cui}},
  \bibinfo{author}{\bibfnamefont{Y.}~\bibnamefont{Wang}}, \bibnamefont{and}
  \bibinfo{author}{\bibfnamefont{F.}~\bibnamefont{Zhou}},
  \bibinfo{journal}{Physical Review A} \textbf{\bibinfo{volume}{78}},
  \bibinfo{pages}{050701} (\bibinfo{year}{2008}).

\bibitem[{\citenamefont{Barnett et~al.}(2010)\citenamefont{Barnett, Sau, and
  Das~Sarma}}]{barnett2010}
\bibinfo{author}{\bibfnamefont{R.}~\bibnamefont{Barnett}},
  \bibinfo{author}{\bibfnamefont{J.~D.} \bibnamefont{Sau}}, \bibnamefont{and}
  \bibinfo{author}{\bibfnamefont{S.}~\bibnamefont{Das~Sarma}},
  \bibinfo{journal}{Physical Review A} \textbf{\bibinfo{volume}{82}},
  \bibinfo{pages}{031602} (\bibinfo{year}{2010}).

\bibitem[{\citenamefont{Zhou}(2001)}]{zhou01}
\bibinfo{author}{\bibfnamefont{F.}~\bibnamefont{Zhou}},
  \bibinfo{journal}{Physical Review Letters} \textbf{\bibinfo{volume}{87}},
  \bibinfo{pages}{080401} (\bibinfo{year}{2001}).

\bibitem[{\citenamefont{Wright et~al.}(2009)\citenamefont{Wright, Leslie,
  Hansen, and Bigelow}}]{wright2009}
\bibinfo{author}{\bibfnamefont{K.~C.} \bibnamefont{Wright}},
  \bibinfo{author}{\bibfnamefont{L.~S.} \bibnamefont{Leslie}},
  \bibinfo{author}{\bibfnamefont{A.}~\bibnamefont{Hansen}}, \bibnamefont{and}
  \bibinfo{author}{\bibfnamefont{N.~P.} \bibnamefont{Bigelow}},
  \bibinfo{journal}{Physical Review Letters} \textbf{\bibinfo{volume}{102}},
  \bibinfo{pages}{030405} (\bibinfo{year}{2009}).

\bibitem[{\citenamefont{Ruostekoski and Anglin}(2003)}]{ruostekoski2003}
\bibinfo{author}{\bibfnamefont{J.}~\bibnamefont{Ruostekoski}} \bibnamefont{and}
  \bibinfo{author}{\bibfnamefont{J.~R.} \bibnamefont{Anglin}},
  \bibinfo{journal}{Physical Review Letters} \textbf{\bibinfo{volume}{91}},
  \bibinfo{pages}{190402} (\bibinfo{year}{2003}).

\bibitem[{\citenamefont{Schmaljohann et~al.}(2004)\citenamefont{Schmaljohann,
  Erhard, Kronjager, Kottke, vanStaa, Cacciapuoti, Arlt, Bongs, and
  Sengstock}}]{schmaljohann04}
\bibinfo{author}{\bibfnamefont{H.}~\bibnamefont{Schmaljohann}},
  \bibinfo{author}{\bibfnamefont{M.}~\bibnamefont{Erhard}},
  \bibinfo{author}{\bibfnamefont{J.}~\bibnamefont{Kronjager}},
  \bibinfo{author}{\bibfnamefont{M.}~\bibnamefont{Kottke}},
  \bibinfo{author}{\bibfnamefont{S.}~\bibnamefont{vanStaa}},
  \bibinfo{author}{\bibfnamefont{L.}~\bibnamefont{Cacciapuoti}},
  \bibinfo{author}{\bibfnamefont{J.~J.} \bibnamefont{Arlt}},
  \bibinfo{author}{\bibfnamefont{K.}~\bibnamefont{Bongs}}, \bibnamefont{and}
  \bibinfo{author}{\bibfnamefont{K.}~\bibnamefont{Sengstock}},
  \bibinfo{journal}{Physical Review Letters} \textbf{\bibinfo{volume}{92}},
  \bibinfo{pages}{040402} (\bibinfo{year}{2004}).

\bibitem[{\citenamefont{Klempt et~al.}(2010)\citenamefont{Klempt, Topic,
  Gebreyesus, Scherer, Henninger, Hyllus, Ertmer, Santos, and
  Arlt}}]{klempt2010}
\bibinfo{author}{\bibfnamefont{C.}~\bibnamefont{Klempt}},
  \bibinfo{author}{\bibfnamefont{O.}~\bibnamefont{Topic}},
  \bibinfo{author}{\bibfnamefont{G.}~\bibnamefont{Gebreyesus}},
  \bibinfo{author}{\bibfnamefont{M.}~\bibnamefont{Scherer}},
  \bibinfo{author}{\bibfnamefont{T.}~\bibnamefont{Henninger}},
  \bibinfo{author}{\bibfnamefont{P.}~\bibnamefont{Hyllus}},
  \bibinfo{author}{\bibfnamefont{W.}~\bibnamefont{Ertmer}},
  \bibinfo{author}{\bibfnamefont{L.}~\bibnamefont{Santos}}, \bibnamefont{and}
  \bibinfo{author}{\bibfnamefont{J.~J.} \bibnamefont{Arlt}},
  \bibinfo{journal}{Physical Review Letters} \textbf{\bibinfo{volume}{104}},
  \bibinfo{pages}{195303} (\bibinfo{year}{2010}).

\bibitem[{\citenamefont{Naik and Raman}(2005)}]{naik05}
\bibinfo{author}{\bibfnamefont{D.~S.} \bibnamefont{Naik}} \bibnamefont{and}
  \bibinfo{author}{\bibfnamefont{C.}~\bibnamefont{Raman}},
  \bibinfo{journal}{Physical Review A (Atomic, Molecular, and Optical Physics)}
  \textbf{\bibinfo{volume}{71}}, \bibinfo{pages}{033617}
  (\bibinfo{year}{2005}).

\bibitem[{\citenamefont{Mewes et~al.}(1997)\citenamefont{Mewes, Andrews, Kurn,
  Durfee, Townsend, and Ketterle}}]{mewe97}
\bibinfo{author}{\bibfnamefont{M.-O.} \bibnamefont{Mewes}},
  \bibinfo{author}{\bibfnamefont{M.~R.} \bibnamefont{Andrews}},
  \bibinfo{author}{\bibfnamefont{D.~M.} \bibnamefont{Kurn}},
  \bibinfo{author}{\bibfnamefont{D.~S.} \bibnamefont{Durfee}},
  \bibinfo{author}{\bibfnamefont{C.~G.} \bibnamefont{Townsend}},
  \bibnamefont{and} \bibinfo{author}{\bibfnamefont{W.}~\bibnamefont{Ketterle}},
  \bibinfo{journal}{Physical Review Letters} \textbf{\bibinfo{volume}{78}},
  \bibinfo{pages}{582} (\bibinfo{year}{1997}).

\bibitem[{\citenamefont{Stamper-Kurn}(1999)}]{stam99thesis}
\bibinfo{author}{\bibfnamefont{D.~M.} \bibnamefont{Stamper-Kurn}},
  \bibinfo{type}{Ph.d. thesis}, \bibinfo{school}{MIT} (\bibinfo{year}{1999}).

\bibitem[{\citenamefont{Saito and Ueda}(2005)}]{saito05}
\bibinfo{author}{\bibfnamefont{H.}~\bibnamefont{Saito}} \bibnamefont{and}
  \bibinfo{author}{\bibfnamefont{M.}~\bibnamefont{Ueda}},
  \bibinfo{journal}{Physical Review A} \textbf{\bibinfo{volume}{72}},
  \bibinfo{pages}{023610} (\bibinfo{year}{2005}).

\bibitem[{\citenamefont{Stapelfeldt and Seideman}(2003)}]{stapelfeldt2003}
\bibinfo{author}{\bibfnamefont{H.}~\bibnamefont{Stapelfeldt}} \bibnamefont{and}
  \bibinfo{author}{\bibfnamefont{T.}~\bibnamefont{Seideman}},
  \bibinfo{journal}{Reviews of Modern Physics} \textbf{\bibinfo{volume}{75}},
  \bibinfo{pages}{543} (\bibinfo{year}{2003}).

\bibitem[{\citenamefont{Scherer et~al.}(2010)\citenamefont{Scherer, Lucke,
  Gebreyesus, Topic, Deuretzbacher, Ertmer, Santos, Arlt, and
  Klempt}}]{scherer2010}
\bibinfo{author}{\bibfnamefont{M.}~\bibnamefont{Scherer}},
  \bibinfo{author}{\bibfnamefont{B.}~\bibnamefont{Lucke}},
  \bibinfo{author}{\bibfnamefont{G.}~\bibnamefont{Gebreyesus}},
  \bibinfo{author}{\bibfnamefont{O.}~\bibnamefont{Topic}},
  \bibinfo{author}{\bibfnamefont{F.}~\bibnamefont{Deuretzbacher}},
  \bibinfo{author}{\bibfnamefont{W.}~\bibnamefont{Ertmer}},
  \bibinfo{author}{\bibfnamefont{L.}~\bibnamefont{Santos}},
  \bibinfo{author}{\bibfnamefont{J.~J.} \bibnamefont{Arlt}}, \bibnamefont{and}
  \bibinfo{author}{\bibfnamefont{C.}~\bibnamefont{Klempt}},
  \bibinfo{journal}{Physical Review Letters} \textbf{\bibinfo{volume}{105}},
  \bibinfo{pages}{135302} (\bibinfo{year}{2010}).

\bibitem[{\citenamefont{McGuirk et~al.}(2003)\citenamefont{McGuirk, Harber,
  Lewandowski, and Cornell}}]{mcguirk03}
\bibinfo{author}{\bibfnamefont{J.~M.} \bibnamefont{McGuirk}},
  \bibinfo{author}{\bibfnamefont{D.~M.} \bibnamefont{Harber}},
  \bibinfo{author}{\bibfnamefont{H.~J.} \bibnamefont{Lewandowski}},
  \bibnamefont{and} \bibinfo{author}{\bibfnamefont{E.~A.}
  \bibnamefont{Cornell}}, \bibinfo{journal}{Physical Review Letters}
  \textbf{\bibinfo{volume}{91}}, \bibinfo{pages}{150402}
  (\bibinfo{year}{2003}).

\bibitem[{\citenamefont{Bray}(1994)}]{bray94}
\bibinfo{author}{\bibfnamefont{A.}~\bibnamefont{Bray}},
  \bibinfo{journal}{Advances in Physics} \textbf{\bibinfo{volume}{43}},
  \bibinfo{pages}{357} (\bibinfo{year}{1994}).

\end{thebibliography}

\end{document}